\documentclass[12pt]{elsart}
\usepackage{amssymb}
\usepackage{epsfig}
\usepackage{graphicx}

\def\Vec#1{{\bf #1}}

\def\D{{\mathrm d}}
\def\E{{\mathrm e}}

\journal{Physics Letters B}

\begin{document}

\begin{frontmatter}
\title{On the $\cos 2 \phi$ asymmetry \\ in unpolarized
leptoproduction}

\author[it]{Vincenzo Barone},
\author[pku]{Zhun Lu} and
\author[pku,it]{Bo-Qiang
Ma} 
\address[it]{Di.S.T.A., Universit\`a del Piemonte Orientale
``A. Avogadro'', \\ and INFN, Gruppo Collegato di Alessandria, 15100
Alessandria, Italy}
\address[pku]{Department of Physics, Peking University, Beijing 100871, China}

\begin{abstract}

We investigate the origin of the $\cos 2 \phi$ azimuthal
asymmetry in unpolarized semiinclusive DIS.
The contributions to this asymmetry arising from the
intrinsic transverse motion
of quarks are explicitly evaluated, and predictions
for the HERMES and COMPASS kinematic regimes are presented. We show that
the effect of the leading-twist
Boer-Mulders function $h_1^{\perp}(x, \Vec k_T^2)$,
which describes a correlation between the transverse
momentum and the transverse spin of quarks,
is quite significant and may also account for a
part of the $\cos 2 \phi$ asymmetry measured
by ZEUS in the perturbative domain.

\end{abstract}
\end{frontmatter}

\vspace{1cm}

\section{Introduction}
\label{intro}

The importance of the transverse-momentum distributions
of quarks for a full understanding of the structure
of hadrons has been widely recognized in the last
decade \cite{levelt,kotzinian,mulders,bm}.
In semi-inclusive deep
inelastic scattering (SIDIS), the $\Vec k_T$-dependent
distributions give rise
to various azimuthal
and/or single-spin asymmetries,
which are currently under direct experimental scrutiny
\cite{Airapetian:2004tw,compass}.
Two leading-twist distributions of great relevance
for their phenomenological implications
are the Sivers function $f_{1T}^{\perp}(x,
\Vec k_T^2)$ \cite{sivers} and its chirally-odd partner
$h_1^{\perp}(x, \Vec k_T^2)$, the so-called Boer--Mulders
function \cite{bm}. These two distributions describe
time-reversal odd correlations between the intrinsic momenta of quarks
and transverse spin vectors \cite{bdr}. In particular,
$f_{1T}^{\perp}$ represents an azimuthal
asymmetry of unpolarized quarks inside a transversely polarized
hadron, whereas $h_1^{\perp}$ represents
a transverse-polarization asymmetry of quarks inside
an unpolarized hadron. Recently,
it has been proven by a direct
calculation \cite{bhs02} that
 $f_{1T}^{\perp}$
and $h_1^{\perp}$ are non-vanishing:
interference diagrams with  a gluon exchanged
between the struck quark and the target remnant generate
non-zero asymmetries.
The presence of a quark transverse momentum smaller than
$Q$ ensures that these asymmetries are proportional to $M/k_T$, rather than to
$M/Q$, and therefore are leading-twist quantities.
Moreover, a careful consideration of the Wilson-line structure
of $\Vec k_T$-dependent
parton densities shows that
 $f_{1T}^{\perp}$
and $h_1^{\perp}$ are not forbidden by time-reversal invariance
\cite{collins02,belitsky} (for a possible chiral origin of
these distributions, see \cite{drago}).

The Sivers function $f_{1T}^{\perp}$ is known
to be responsible for a $\sin (\phi - \phi_S)$
 single-spin asymmetry in transversely polarized SIDIS
\cite{Airapetian:2004tw,compass,anselmino05}.
The Boer-Mulders function $h_1^{\perp}$
produces azimuthal asymmetries in {\em unpolarized}
reactions. Boer \cite{boer} argued that
it can account for the observed $ \cos 2 \phi$
asymmetries in unpolarized $\pi N$ Drell-Yan
processes \cite{na10,conway}.
This was quantitatively confirmed in \cite{lm04,lm05},
where $h_1^{\perp}$ was calculated in a simple
quark-spectator model and shown to explain
the Drell-Yan data fairly well.

A similar $\cos 2 \phi$ asymmetry occurs in unpolarized
leptoproduction. As we shall see, there are three possible
mechanisms generating this asymmetry: 1) non-collinear
kinematics at order $k_T^2/Q^2$ \cite{cahn}; 2) the leading-twist
Boer-Mulders function \cite{bm} coupling
to a specular fragmentation function,
the so-called Collins function \cite{collins93},
which describes the fragmentation of transversely
polarized quarks into unpolarized hadrons; 3) perturbative gluon radiation
\cite{georgi,mendez,konig,chay}.
The purpose of this paper is to study the first two sources
of the $ \cos 2 \phi$ asymmetry, both related to the intrinsic
transverse motion of quarks. They are especially relevant
in the HERMES kinematic regime ($\langle Q^2 \rangle
\sim 2$ GeV$^2$), but the Boer-Mulders contribution, being
leading twist,
can also survive at higher $Q^2$ and partly account for the
asymmetry measured by ZEUS in this domain \cite{zeus}.

In recent years, the $ \cos 2 \phi$ asymmetry in leptoproduction was
phenomenologically studied by
some authors \cite{oga,gg03}.
In \cite{oga} only the
$\mathcal{O}(k_T^2/Q^2)$ term and the perturbative
contribution were included, whereas the Boer-Mulders
effect was not considered. Our calculation is more similar
to that presented in \cite{gg03}, the main differences
being that we use a model
for $h_1^{\perp}$ adjusted on the
Drell-Yan data \cite{lm05},
 and compute the asymmetry according to
its experimental definition (which incorporates a
cutoff on the transverse momentum of the final hadron).

\section{The $\cos 2 \phi$ asymmetry in unpolarized SIDIS}

\begin{figure}[t]
\begin{center}
\scalebox{0.75}{\includegraphics{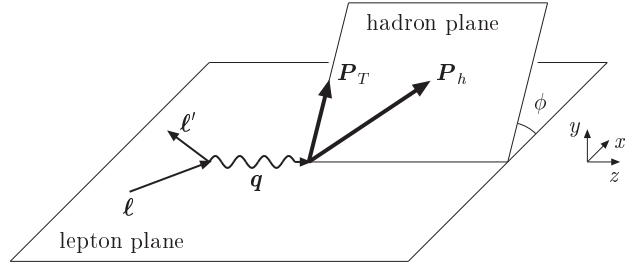}}
\caption{\small Lepton and hadron planes in semi-inclusive 
deep inelastic scattering.}
\label{plane}
\end{center}
\end{figure}

The process we are interested in is unpolarized SIDIS:
\begin{equation}
l (\ell) \, + \, p (P) \, \rightarrow \, l' (\ell')
\, + \, h (P_h) \, + \, X (P_X)\,.
\label{sidis}
\end{equation}
The SIDIS cross section is expressed in terms of
the invariants
\begin{equation}
x = \frac{Q^2}{2 \, P \cdot q}, \;\;\;
y =  \frac{P \cdot q}{P \cdot \ell} ,
\;\;\;
z = \frac{P \cdot P_h}{P \cdot q}\,,
\end{equation}
where $ q = \ell - \ell'$ and $Q^2 \equiv - q^2$.
We adopt a reference frame such that
the virtual photon and the target proton are collinear
and directed along the $z$ axis, with the 
photon moving in the positive $z$ direction 
(Fig.~\ref{plane}). We denote by $\Vec k_T$ the transverse
momentum of the quark inside the proton, and by $\Vec P_T$ the
transverse momentum of the hadron $h$. The transverse momentum
of $h$ with respect to the direction of the fragmenting
quark will be called $\Vec p_T$. All azimuthal angles
are referred to the lepton scattering plane
(we call $\phi$ the azimuthal angle of the hadron $h$, 
see Fig.~\ref{plane}).

Taking the intrinsic motion of quarks into account, the
SIDIS cross section reads at leading order
\begin{eqnarray}
& &  \frac{\D \sigma}{\D x \, \D y \, \D z \, \D^2 \Vec P_T}
 =  \frac{2 \pi \alpha_{\rm em}^2 s}{Q^4} \, \sum_a
e_a^2 \, x [1 + (1 - y)^2] \nonumber \\
& &  \hspace{1cm} \times \, \int \D^2 \Vec k_T \,
\int \D^2 \Vec p_T \,
\delta^2 (\Vec P_T - z \Vec k_T - \Vec p_T)
\, f_1^a (x, \Vec k_T^2) \,
D_1^a (z, \Vec p_T^2) \,,
\label{cross1}
\end{eqnarray}
where $f_1^a (x, \Vec k_T^2)$ is the unintegrated
number density of quarks of flavor $a$ and
$D_1^a (z, \Vec p_T^2)$ is the transverse-momentum
dependent fragmentation function of quark $a$ into
the final hadron. We recall that the non-collinear factorization
theorem for SIDIS has been recently proven
by Ji, Ma and Yuan \cite{Ji:2004xq} for
$P_{T} \ll Q$.

As shown long time ago by Cahn \cite{cahn}, the transverse-momentum
kinematics generates a $\cos 2 \phi$ contribution
to the unpolarized SIDIS cross section, which
has the form
\begin{eqnarray}
& & \left. \frac{\D \sigma^{(HT)}}{\D x \, \D y \, \D z \, \D^2 \Vec P_T}
\right \vert_{\cos 2 \phi}
 =  \frac{8 \pi \alpha_{\rm em}^2 s}{Q^4} \, \sum_a
e_a^2 \, x (1 - y) \nonumber \\
& & \hspace{2cm} \times \,
 \int \D^2 \Vec k_T \, \int \D^2 \Vec p_T
\,\delta^2 (\Vec P_T - z \Vec k_T - \Vec p_T)
\nonumber \\
& & \hspace{2cm} \times \,
\frac{2 \, (\Vec k_T \cdot \Vec h)^2 - \Vec k_T^2}{Q^2} \,
f_1^a (x, \Vec k_T^2) \,
D_1^a (z, \Vec p_T^2)\, \cos 2 \phi \,,
\label{cross2}
\end{eqnarray}
 where $\Vec h \equiv \Vec P_T/P_T$. Notice that this contribution
is of order $k_T^2/Q^2$, hence it is a (kinematic) higher twist effect.

The second $k_T$-dependent source of the $\cos 2 \phi$
asymmetry involves the Boer-Mulders distribution $h_1^{\perp}$
coupled to the Collins fragmentation function $H_1^{\perp}$
of the produced hadron. The explicit expression
of this contribution
to the cross section is \cite{bm}
\begin{eqnarray}
& & \left. \frac{\D \sigma^{(LT)}}{\D x \, \D y \, \D z \, \D^2 \Vec P_T}
\right \vert_{\cos 2 \phi}
 =  \frac{4 \pi \alpha_{\rm em}^2 s}{Q^4} \, \sum_a
e_a^2 \, x (1 - y) \nonumber \\
& & \hspace{2cm} \times \,
 \int \D^2 \Vec k_T \, \int \D^2 \Vec p_T
\,\delta^2 (\Vec P_T - z \Vec k_T - \Vec p_T)
\nonumber \\
& & \hspace{2cm} \times \,
\frac{2 \, \Vec h \cdot \Vec k_T \,
\Vec h \cdot \Vec p_T - \Vec k_T \cdot \Vec p_T}{z  M M_h} \,
h_1^{\perp a} (x, \Vec k_T^2) \,
H_1^{\perp a} (z, \Vec p_T^2)\, \cos 2 \phi \,,
\label{cross3}
\end{eqnarray}
It should be noticed that this is a leading-twist contribution, not
suppressed by inverse powers of $Q$.

The asymmetry measured in experiments is defined as
\begin{equation}
\langle \cos 2 \phi \rangle =
\frac{\int \D \sigma \, \cos 2 \phi}{\int \D \sigma}\,,
\end{equation}
where the integrations are performed over the measured
ranges of $x, y, z$ and with a lower cutoff $P_c$ on $P_T$, 
which is the minimum value of $P_T$ of the detected charged particles.
Using Eqs.~(\ref{cross1}) and (\ref{cross3}),
$\langle \cos 2 \phi_h \rangle$ is given by
\begin{equation}
\langle \cos 2 \phi \rangle = \frac{\int \int \int \int \,
 \sum_a e_a^2 \, 2 x (1- y) \, \{ \mathcal{A} [f_1^a, D_1^a] +
\frac{1}{2} \, \mathcal{B} [h_1^{\perp a}, H_1^{\perp a}]
\} }{\int \int \int \int \,
\sum_a e_a^2 \, x   [1 +
(1 - y)^2]  \, \mathcal{C} [f_1^a,
D_1^a]}\,,
\label{asymmetry}
\end{equation}
where
\begin{equation}
\int \int \int \int \equiv
\int_{P_c}^{P_{T, max}} \D P_T \, P_T
\, \int_{x_1}^{x_2} \D x \,
\int_{y_1}^{y_2} \D y \, \int_{z_1}^{z_2} \D z \,
\end{equation}
and
($\chi$ is the angle between $\Vec P_T$
and $\Vec k_T$)
\begin{eqnarray}
\mathcal{A} [f_1^{a}, D_1^{a}] &\equiv&
 \int \D^2 \Vec k_T \, \int \D^2 \Vec p_T
\,\delta^2 (\Vec P_T - z \Vec k_T - \Vec p_T)
\nonumber \\
& &  \times \,
\frac{2 \, (\Vec k_T \cdot \Vec h)^2 - \Vec k_T^2}{Q^2} \,
f_1^a (x, \Vec k_T^2) \,
D_1^a (z, \Vec p_T^2)\, \cos 2 \phi
\nonumber \\
& =&
\int_0^{\infty} \D k_T \, k_T \, \int_0^{2 \pi} \D \chi
\, \frac{2 \Vec k_T^2 \, \cos^2 \chi - \Vec k_T^2}{Q^2}
\nonumber \\
& & \times \,  f_1^{a}(x, \Vec k_T^2)
\, D_1^{a}(z, \vert \Vec P_T - z \Vec k_T \vert^2)\,,
\label{convol1}
\end{eqnarray}
\begin{eqnarray}
\mathcal{B} [h_1^{\perp a}, H_1^{\perp a}] &\equiv&
 \int \D^2 \Vec k_T \, \int \D^2 \Vec p_T
\,\delta^2 (\Vec P_T - z \Vec k_T - \Vec p_T)
\nonumber \\
& &  \times \,
\frac{2 \, \Vec h \cdot \Vec k_T \,
\Vec h \cdot \Vec p_T - \Vec k_T \cdot \Vec p_T}{z  M M_h} \,
h_1^{\perp a} (x, \Vec k_T^2) \,
H_1^{\perp a} (z, \Vec p_T^2)
\nonumber \\
& =&
\int_0^{\infty} \D k_T \, k_T \, \int_0^{2 \pi} \D \chi
\, \frac{\Vec k_T^2 + (P_T/z)\, k_T
\, \cos \chi - 2 \, \Vec k_T^2 \, \cos^2 \chi}{M M_h}
\nonumber \\
& & \times \, h_1^{\perp a}(x, \Vec k_T^2)
\, H_1^{\perp a}(z, \vert \Vec P_T - z \Vec k_T \vert^2)\,,
\label{convol2}
\end{eqnarray}
\begin{eqnarray}
\mathcal{C} [f_1^{a}, D_1^{a}] &\equiv&
\int \D^2 \Vec k_T \,
\int \D^2 \Vec p_T \,
\delta^2 (\Vec P_T - z \Vec k_T - \Vec p_T)
\, f_1^a (x, \Vec k_T^2) \,
D_1^a (z, \Vec p_T^2)
\nonumber \\
& =&
\int_0^{\infty} \D k_T \, k_T \, \int_0^{2 \pi} \D \chi
\,  f_1^{a}(x, \Vec k_T^2)
\, D_1^{a}(z, \vert \Vec P_T - z \Vec k_T \vert^2)\,.
\label{convol3}
\end{eqnarray}

\section{Calculation and Results}

In order to calculate $\langle \cos 2 \phi \rangle$
one needs to know the $k_T$- and $p_T$-dependent
distribution and fragmentation functions appearing
in Eqs.~(\ref{convol1})-(\ref{convol3}).
Independent information on the Boer-Mulders function
$h_1^{\perp}(x, \Vec k_T^2)$ can be obtained from
the study of the $\cos 2 \phi$ azimuthal asymmetry
in unpolarized Drell-Yan processes, which has been
measured in $\pi N$ collisions
\cite{na10,conway}.
In \cite{lm04,lm05} this asymmetry was estimated
by computing the $h_1^{\perp}$ distribution of the
pion and of the nucleon in a quark spectator model
\cite{jmr97,bsy04}. To compute
the $\cos 2 \phi$ azimuthal asymmetry in SIDIS
we adopt the same distributions
$h_1^{\perp} (x, \Vec k_T^2)$ and
$f_1 (x, \Vec k_T^2)$ used in \cite{lm05}.
We assume that the observables are dominated by
$u$ quarks (i.e., we consider $\pi^+$ production).
The set of the
transverse-momentum dependent distribution functions
is (for simplicity, we consider a spectator scalar diquark
\cite{bsy04,lm05})
\begin{eqnarray}
& & f_{1}^{u}(x,\Vec k_T^2) = N \, (1 - x)^3 \, \frac{(x
M + m)^2 + \Vec k_T^2}{(L^2 + \Vec k_T^2)^4},
\label{f1} \\
& & h_{1}^{\perp u}(x, \Vec k_T^2) = \frac{4}{3}
\,  \alpha_s \, N\,  (1 - x)^{3} \,
\frac{M \, (x M + m)}{[L^2 \, (L^2 +
\Vec k_T^2)^3]},
\label{h1perp}
\end{eqnarray}
where $N$ is a normalization constant, $m$ is the
constituent quark mass, and
\begin{equation}
L^2=(1 - x) \, \Lambda^2 + x \, M_d^2  - x \, (1 - x) \, M^2\,.
\end{equation}
Here $\Lambda$ is a cutoff appearing in the
nucleon-quark-diquark vertex and
$M_d$ is the mass of the scalar diquark.
As it is typical of all model calculations of
quark distribution functions, we expect that Eqs.~(\ref{f1}) and
(\ref{h1perp})
should be valid at low $Q^2$ values, of order of 1 GeV$^2$.
The average transverse momentum of quarks inside the
target, as computed from (\ref{f1}), turns out to be
$\langle k_T^2 \rangle^{1/2} \simeq 0.54$ GeV.

\begin{figure}[t]
\begin{center}
\scalebox{0.75}{\includegraphics{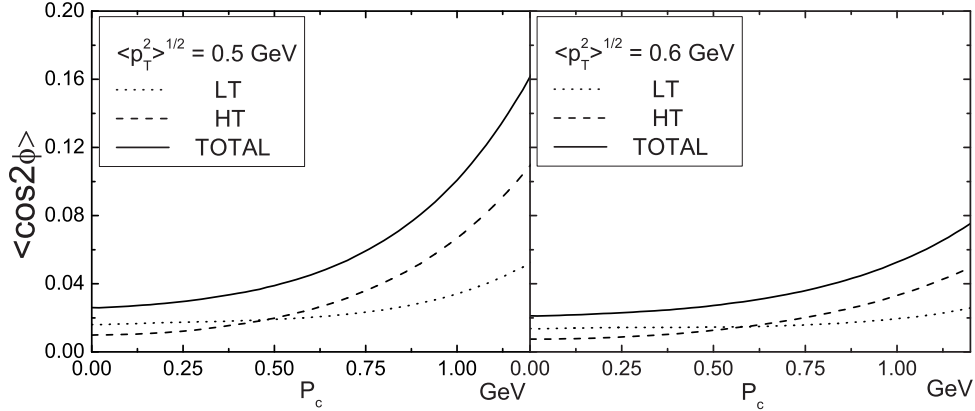}}
\caption{\small The
SIDIS $\cos 2 \phi$ azimuthal asymmetry in the HERMES domain
as a function of the cutoff $P_c$, for two values of $\langle p_T^2
\rangle^{1/2}$.
The dotted curve is the leading-twist Boer-Mulders contribution,
the dashed curve is the higher-twist term,
the solid curve is the sum of the two contributions.}
\label{cos2phi_hermes}
\end{center}
\end{figure}

Coming to the fragmentation functions, for $H_1^{\perp}$
we adopt the simple parametrization suggested by
Collins \cite{collins93}
\begin{equation}
\frac{H_1^{\perp}(z, \Vec p_T^2)}{D_1(z, \Vec p_T^2)}=\frac{M_C
M_h}{M_C^2 + \Vec p_T^2/z^2},
\label{cff}
\end{equation}
where $M_C$ is a free parameter. We assume a
Gaussian dependence for the unintegrated unpolarized
fragmentation function:
\begin{equation}
D_1(z, \Vec p_T^2)=D_1 (z) \frac{1}{\pi \langle p_T^2 \rangle}
\, \E^{- \Vec p_T^2/\langle p_T^2 \rangle}\,,
\label{d1}
\end{equation}
so that $\int \D^2 \Vec p_T \, D_1 (z, \Vec p_T^2) =
D_1 (z)$. Finally,  the integrated unpolarized
fragmentation function for pions $D_1(z)$ is taken from
the Kretzer--Leader--Christova parametrization~\cite{Kre01},
\begin{equation}
D_1 (z) = 0.689 \, z^{-1.039} \, (1 - z)^{1.241}
\end{equation}
valid at $\langle Q^2 \rangle = 2.5$ GeV$^2$.
For the parameters in Eqs.~(\ref{f1}) and (\ref{h1perp}) we
choose the values $M_d=0.8$ GeV, $m=0.3$ GeV,
$\Lambda=0.6$ GeV, $\alpha_s=0.3$, which are the same
as in \cite{lm05}.  As for the parameters in
Eqs.~(\ref{cff}) and (\ref{d1}), we fix $M_C$ to 0.3 GeV and
show results for two values of the average
transverse momentum: $\langle p_T^2
\rangle^{1/2}=0.5$ GeV and $0.6$ GeV (we checked that a variation of $M_C$
is reproduced by a change of $\langle p_T^2
\rangle^{1/2}$).

\begin{figure}
\begin{center}
\scalebox{0.75}{\hspace{-0.5cm}\includegraphics{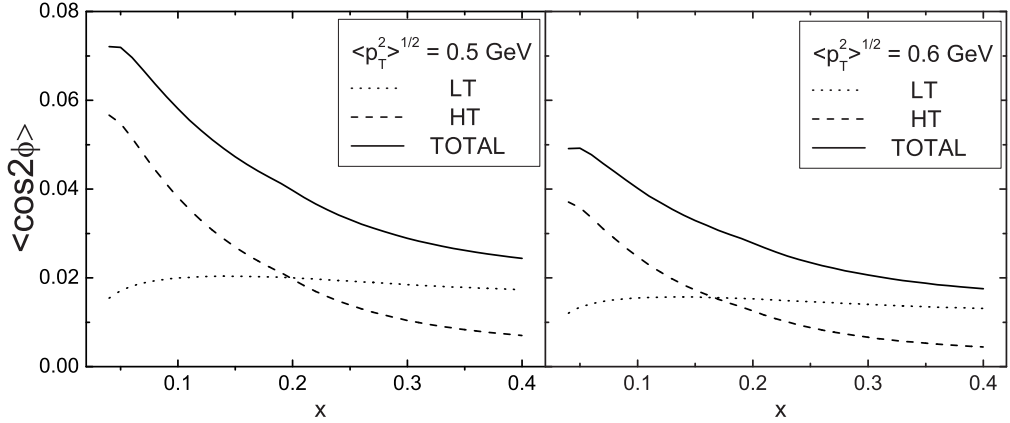}}
\caption{\small The
SIDIS $\cos 2 \phi$ azimuthal asymmetry in the HERMES domain,
as a function of $x$ with $P_c = 0.5$ GeV.
The dotted curve is the leading-twist Boer-Mulders contribution,
the dashed curve is the higher-twist term,
the solid curve is the sum of the two contributions.}
\label{cos2phi_hermesx}
\end{center}
\end{figure}

\begin{figure}
\begin{center}
\scalebox{0.75}{\hspace{-0.5cm}\includegraphics{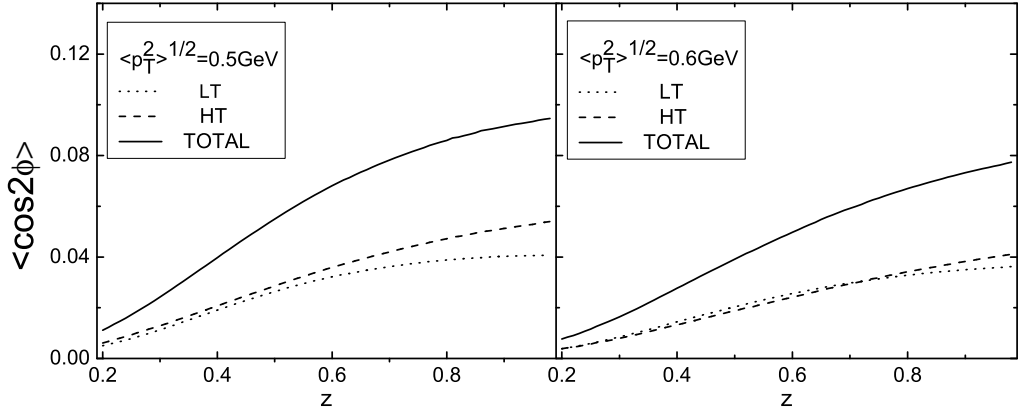}}
\caption{\small The
SIDIS $\cos 2 \phi$ azimuthal asymmetry in the HERMES domain,
as a function of $z$ with $P_c = 0.5$ GeV.
The dotted curve is the leading-twist Boer-Mulders contribution,
the dashed curve is the higher-twist term,
the solid curve is the sum of the two contributions.}
\label{cos2phi_hermesz}
\end{center}
\end{figure}

The HERMES kinematics is characterized by the following
ranges:
$0.02 < x < 0.4$, $0.1 < y < 0.85$, $0.2 < z < 1$,
$\langle Q^2 \rangle = 2$ GeV$^2$. Our predictions
for the $\cos 2 \phi$ asymmetry in this regime
are displayed in Fig.~\ref{cos2phi_hermes}, where we
show separately the higher-twist term and the
leading-twist Boer-Mulders
contribution. For a typical transverse momentum cutoff $P_c
=0.5$ GeV, these two terms are comparable and
the predicted asymmetry lies in the range $\langle \cos 2 \phi
\rangle = 0.02-0.04$. The $x$-dependence (with $z$ integrated over
the accessible interval) and the $z$-dependence (with $x$
integrated over the accessible interval) are shown
in Figs.~\ref{cos2phi_hermesx} and \ref{cos2phi_hermesz},
respectively. As one can see, the asymmetry is larger
at small $x$ and large $z$. 

\begin{figure}
\begin{center}
\scalebox{0.75}{\hspace{-0.5cm}\includegraphics{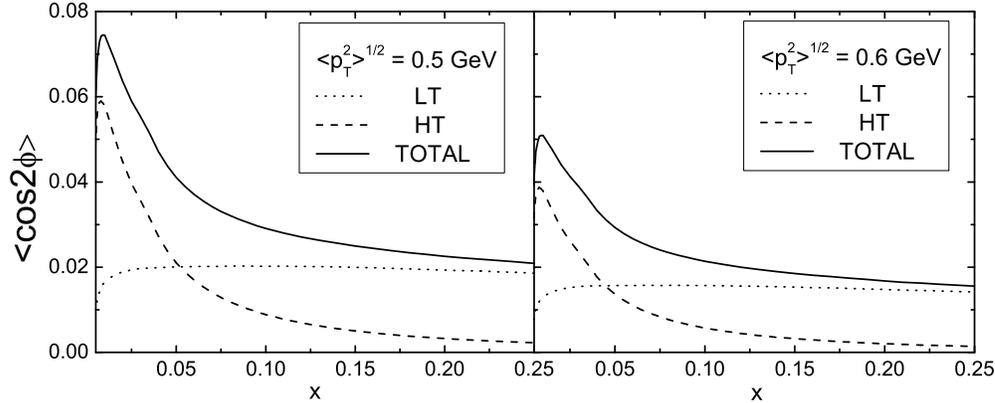}}
\caption{\small The
SIDIS $\cos 2 \phi$ azimuthal asymmetry in the COMPASS domain,
as a function of $x$ with $P_c = 0.5$ GeV.
The dotted curve is the leading-twist Boer-Mulders contribution,
the dashed curve is the higher-twist term,
the solid curve is the sum of the two contributions.}
\label{cos2phi_compassx}
\end{center}
\end{figure}

In Fig.~\ref{cos2phi_compassx} we plot our results for the 
$x$-dependent asymmetry (integrated over $z$) 
in the COMPASS kinematic domain. The correlation between 
$x$ and $Q^2$ is such that the lowest $x$ bin ($x = 0.005$) corresponds 
to $Q^2 \approx 1$ GeV$^2$, whereas the highest $x$ bin in 
Fig.~\ref{cos2phi_compassx} ($x = 0.25$) 
corresponds to $Q^2 \approx 24$ GeV$^2$. Again, the asymmetry 
is of order of few percent and decreases with $x$.

\begin{figure}
\begin{center}
\scalebox{0.8}{\includegraphics{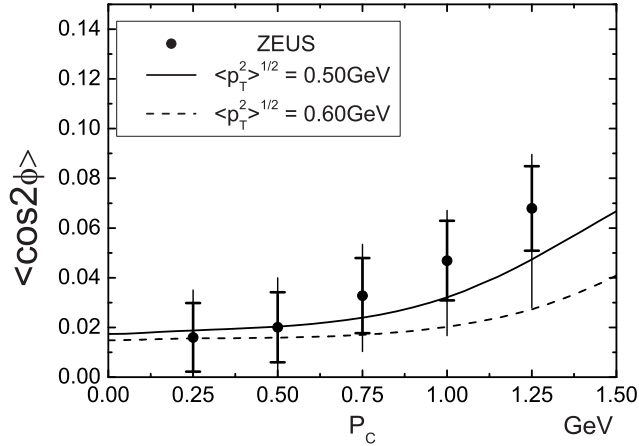}}\caption{\small The
SIDIS  $\cos 2 \phi$ azimuthal asymmetry as a function
of the cutoff $P_c$ in the ZEUS domain. Data
are from \cite{zeus}.}
\label{cos2phi_zeus}
\end{center}
\end{figure}

There are available data on the $\cos 2 \phi$ asymmetry
in SIDIS coming from the ZEUS experiment \cite{zeus}.
The ZEUS kinematic ranges are:
$0.01<x<0.1$, $0.2<y<0.8$, $0.2<z<1$, $Q^2 > 180$ GeV$^2$.
At such large $Q^2$ values, the higher twist contribution
is clearly irrelevant. Since only the 
$Q^2$ evolution of the $k_T$ moments of $h_1^{\perp}$
is known \cite{henneman}, and not that of 
$h_1^{\perp}$ itself,  
we assume for simplicity that the distributions (\ref{f1}) and (\ref{h1perp})
scale exactly, i.e. that they are valid for any $Q^2$ (one should 
recall however that Sudakov form factors arising from 
soft gluon contributions may reduce 
the Boer--Mulders asymmetry at very high $Q^2$ \cite{boer2}).
The result for the $\cos 2 \phi$ asymmetry
in the ZEUS kinematic domain is shown in Fig.~\ref{cos2phi_zeus},
where it is compared with the experimental data.
The agreement is rather good for low values of the
$P_T$ cutoff (up to 0.5 GeV). For larger $P_T$ values
one expects of course a relevant perturbative contribution.
Including this contribution is beyond the purpose of this paper,
which is primarily devoted to predictions for the
low-$Q^2$ domain. A more extended analysis of the
$\cos 2 \phi$ asymmetries, taking into account
also the perturbative term, is in progress and will be reported
soon \cite{bmp}.
 
\begin{figure}
\begin{center}
\scalebox{0.8}{\includegraphics{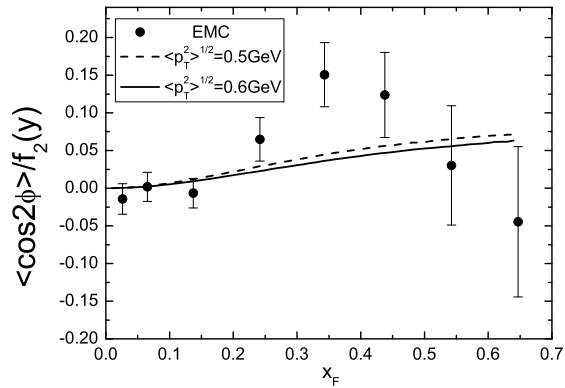}}\caption{\small The
 $\cos 2 \phi$ azimuthal asymmetry (divided by 
$f_2 (y) = (1 - y)/[1 + (1 - y)^2 ]$) as a function
of $x_F \equiv 2 P_L/W$ as measured by EMC \cite{emc}. The curves 
are our predictions.}
\label{cos2phi_emc}
\end{center}
\end{figure}

For completeness we recall 
that long time ago the European Muon Collaboration at CERN  measured 
$\langle \cos 2 \phi \rangle$ for $Q^2 > 4$ GeV$^2$ \cite{emc}. 
The EMC data, however, are affected by large uncertainties 
and do not allow drawing definite conclusions about the 
magnitude and the shape of the asymmetry. The comparison 
of our predictions with these data is shown in Fig.~\ref{cos2phi_emc}.  

In conclusion, we predicted the $\cos 2 \phi$ asymmetry for
semi-inclusive deep inelastic scattering in the kinematic regions
of the HERMES and COMPASS experiments. We found that $\langle \cos 2 \phi
\rangle$ is of order of few percent and tends to be larger in the
small-$x$ and large-$z$ region. The combined analysis of the
future data on $\langle \cos 2 \phi \rangle$ and of the
previous ZEUS measurements in the high-$Q^2$ domain (where higher
twist effects are irrelevant) will allow to get information on the
Boer-Mulders function, shedding light on the correlations between
transverse spin and transverse momenta of quarks.

{\bf Acknowledgements.}
We are grateful to Alexei Prokudin and 
Franco Brada\-mante for useful discussions.
This work is partially supported by the National Natural Science
Foundation of China (No.~10421003), by the Key Grant Project of
the Chinese Ministry of Education (No.~305001), and by the Italian
Ministry of Education, University and Research (PRIN 2003).


\begin{thebibliography}{99}

\bibitem{levelt}
J.~Levelt, P.J.~Mulders, Phys. Rev. {\bf D49} (1994) 96.

\bibitem{kotzinian}
A.~Kotzinian, Nucl. Phys. {\bf B441} (1995) 234.

\bibitem{mulders}
P.J.~Mulders, R.D.~Tangerman, Nucl. Phys.
{\bf B461} (1996) 197.

\bibitem{bm}
D.~Boer, P.J.~Mulders, Phys. Rev. {\bf D57} (1998) 5780.

\bibitem{Airapetian:2004tw}
HERMES Collaboration, A.~Airapetian, et al., Phys. Rev. Lett. {\bf
94} (2005) 012002.

\bibitem{compass}
COMPASS Collaboration, V.Yu.~Alexakhin, et al.,  Phys. Rev. Lett. {\bf
94} (2005) 202002.

\bibitem{sivers} D. Sivers, Phys. Rev. {\bf D41} (1990) 83;
Phys. Rev. {\bf D43} (1991) 261.

\bibitem{bdr}
For a review on transverse polarization phenomena, see V.~Barone,
A.~Drago, P.G.~Ratcliffe, Phys. Rep. {\bf 359} (2002) 1.


\bibitem{bhs02} S.J.~Brodsky, D.S.~Hwang,
I.~Schmidt, Phys. Lett. {\bf B530} (2002) 99;
Nucl. Phys. {\bf B642} (2002) 344.

\bibitem{collins02} J.C. Collins,  Phys. Lett.  {\bf B536} (2002) 43.

\bibitem{belitsky}
A.V.~Belitsky, X.~Ji, F.~Yuan, Nucl. Phys.
{\bf B656} (2003) 165.


\bibitem{drago} M.~Anselmino, A.~Drago, F.~Murgia,
hep-ph/9703303. M.~Anselmino, V.~Barone,
A.~Drago, F.~Murgia, hep-ph/0209073.
A.~Drago, Phys. Rev. {\bf D71} (2005) 057501.



\bibitem{anselmino05}
M.~Anselmino, et al., Phys. Rev. {\bf D71} (2005) 074006.

\bibitem{boer}
D.~Boer, Phys. Rev. {\bf D60} (1999) 014012.

\bibitem{na10} NA10 Collaboration, S.~Falciano, et al.,
Z. Phys. {\bf C31} (1986) 513;\\
NA10 Collaboration, M.~Guanziroli, et al., Z. Phys. {\bf C37}
(1988) 545.

\bibitem{conway} E615 Collaboration, J.S.~Conway, et al.,
Phys. Rev. {\bf D39} (1989) 92.

\bibitem{lm04} Z.~Lu, B.-Q.~Ma, Phys. Rev.  {\bf D70} (2004)
094044;

\bibitem{lm05} Z.~Lu, B.-Q.~Ma,
Phys. Lett. {\bf B615} (2005) 200 (hep-ph/0504184).

\bibitem{cahn}
R.N.~Cahn, Phys. Lett. {\bf B78} (1978) 269;
Phys. Rev. {\bf D40} (1989) 3107.


\bibitem{collins93} J.C.~Collins, Nucl. Phys. {\bf B396} (1993) 161.

\bibitem{georgi}
H.~Georgi, H.D.~Politzer, Phys. Rev. Lett. {\bf 40} (1978) 3.

\bibitem{mendez}
A.~Mendez, Nucl. Phys. {\bf B145} (1978) 199.

\bibitem{konig}
A.~K{\"o}nig, P.~Kroll, Z. Phys. {\bf C16} (1982) 89.

\bibitem{chay}
J.~Chay, S.D.~Ellis, W.J.~Stirling, Phys. Rev.
{\bf D45} (1992) 46.

\bibitem{zeus} ZEUS Collaboration, J. Breitweg, et al.,
Phys. Lett. {\bf B481} (2000) 199.

\bibitem{oga}
K.A.~Oganessyan, H.R.~Avakian, N.~Bianchi, P.~Di Nezza,
Eur. Phys. J. {\bf C5} (1998) 681.

\bibitem{gg03} L.P.~Gamberg, G.R.~Goldstein,
K.A.~Oganessyan, Phys. Rev. {\bf D67} (2003) 071504. L.~Gamberg,
hep-ph/0412367.


\bibitem{Ji:2004xq}
X.~Ji, J.P.~Ma, F.~Yuan,
Phys. Lett. {\bf B597} (2004) 299.

\bibitem{jmr97} R.~Jakob, P.J.~Mulders, J.~Rodrigues, Nucl. Phys.
{\bf A626} (1997) 937.


\bibitem{bsy04} A.~Bacchetta, A.~Sch\"{a}fer,
J.-J.~Yang, Phys. Lett. {\bf B578} (2004) 109.



\bibitem{Kre01}
S.~Kretzer, E.~Leader, E.~Christova, Eur. Phys. J. {\bf C22} (2001)
269.


\bibitem{henneman} 
A.A.~Henneman, D.~Boer, P.J.~Mulders, Nucl. Phys. {\bf B620} (2002) 331. 

\bibitem{boer2}
D.~Boer, Nucl. Phys. {\bf B603} (2001) 195. 

\bibitem{bmp}
V.~Barone, B.-Q.~Ma, A.~Prokudin, work in preparation.

\bibitem{emc}
EMC, M.~Arneodo, et al., Z. Phys. {\bf C34} (1987) 277.



\end{thebibliography}
\end{document}